\documentclass[12pt,a4paper]{article}
\pdfoutput=1
\usepackage{color}
\usepackage{amssymb,amsmath,bm,bbold}
\usepackage{epsf}
\usepackage{epsfig}
\usepackage{relsize}
\usepackage[dvipsnames]{xcolor}
\usepackage[linktoc=page,bookmarks=false,colorlinks=false,linkbordercolor=RoyalBlue,citebordercolor=ForestGreen,urlbordercolor=CornflowerBlue]{hyperref}
\usepackage{latexsym,mathrsfs,dsfont}
\usepackage[normalem]{ulem} 
\usepackage[compress]{cite}
\usepackage{graphicx}
\usepackage{url}
\usepackage{booktabs}
\usepackage{float}
\usepackage{multirow}
\usepackage{changepage}
\usepackage[hypcap]{caption, subcaption}

\usepackage[titles]{tocloft}

\usepackage{tabularx,colortbl}

\usepackage{fancyhdr}
\advance \headheight by 3.0truept       

\allowdisplaybreaks[1]

\definecolor{red}{cmyk}{0,1,1,0.4}
\definecolor{darkgreen}{rgb}{0.0,0.6,0.0}
\definecolor{cDarkGrey}{RGB}{91,91,91}
\definecolor{cGrey}{RGB}{245,243,238}
\definecolor{cBlue}{RGB}{0,110,191}
\definecolor{cLightBlue}{RGB}{214,237,252}
\definecolor{cRed}{RGB}{196,0,100}
\definecolor{cLightRed}{RGB}{254,222,237}
\definecolor{cGreen}{RGB}{0,166,80}
\definecolor{cLightGreen}{RGB}{254,222,237}
\definecolor{cOrange}{RGB}{221,74,44}
\definecolor{cLightOrange}{RGB}{255,215,210}
\definecolor{cPurple}{RGB}{93,35,125}
\definecolor{cLightPurple}{RGB}{241,230,252}
\definecolor{cYellow}{RGB}{252,191,10}
\definecolor{cISSRBlue}{RGB}{0,111,174}
\definecolor{cISSRGrey}{RGB}{167,169,172}

\newcommand{\beq}{\begin{equation}}
\newcommand{\eeq}{\end{equation}}
\newcommand{\be}{\begin{equation}}
\newcommand{\ee}{\end{equation}}
\newcommand{\bi}{\begin{itemize}}
\newcommand{\ei}{\end{itemize}}
\newcommand{\ba}{\begin{array}}
\newcommand{\ea}{\end{array}}
\newcommand{\beqa}{\begin{eqnarray}}
\newcommand{\eeqa}{\end{eqnarray}}
\newcommand{\bea}{\begin{eqnarray}}
\newcommand{\eea}{\end{eqnarray}}
\newcommand{\beqn}{\begin{eqnarray}}
\newcommand{\eeqn}{\end{eqnarray}}

\newcounter{TODO}

\newcommand{\mev}{\text{MeV}}
\newcommand{\GeV}{\,\text{GeV}}

\newcommand{\vcb}{|V_{cb}|}
\newcommand{\vtd}{|V_{td}|}
\newcommand{\vub}{|V_{ub}|}
\newcommand{\vts}{|V_{ts}|}
\newcommand{\vus}{|V_{us}|}

\def\kpn{K^+\rightarrow\pi^+\nu\bar\nu}
\def\klpn{K_{L}\rightarrow\pi^0\nu\bar\nu}
\def\ksm{K_S\to\mu^+\mu^-}

\newcommand{\IM}{\rm{Im}}

\newcommand{\Nf}{N_f}

\newcommand{\eps}{\epsilon}

\begin{document}

\begin{flushleft}
\end{flushleft}

\vspace{-14mm}
\begin{flushright}
  AJB-22-6\\
  TUM-HEP-1393/22
\end{flushright}

\medskip

\begin{center}
{\Large\bf\boldmath
  The Exclusive Vision of Rare $K$ and $B$ Decays\\ and of the Quark Mixing in the
  Standard Model
}
\\[1.0cm]
{\bf
    Andrzej~J.~Buras$^{a,b}$ and
  Elena Venturini$^{b}$
}\\[0.3cm]

{\small
$^a$TUM Institute for Advanced Study,
    Lichtenbergstr. 2a, D-85747 Garching, Germany \\[0.2cm]
$^b$Physik Department, TU M\"unchen, James-Franck-Stra{\ss}e, D-85748 Garching, Germany
}
\end{center}

\vskip 0.5cm

\begin{abstract}
  \noindent
  The most common predictions for rare $K$ and $B$ decay branching ratios  in the Standard Model in the literature are based on   the CKM elements $|V_{cb}|$ and  $|V_{ub}|$  resulting from global fits, that are in the ballpark of their  inclusive and exclusive determinations, respectively. In the present paper we follow another route,  which to our knowledge has not been explored for
  $\Delta M_{s,d}$ and rare $K$ and $B$ decays by anybody to date. We assume, in contrast to the prevailing {\em inclusive} expectations
  for  $|V_{cb}|$,
  that the future true  values of $|V_{cb}|$ and  $|V_{ub}|$  will be both from
  {\em exclusive} determinations; {in practice we use the most recent averages from FLAG.}
   With the precisely known $\vus$ the
  resulting rare decay branching ratios, $\varepsilon_K$, $\Delta M_d$, $\Delta M_s$  and $S_{\psi K_S}$ depend then only on the angles $\beta$ and $\gamma$
  in the unitarity triangle that moreover are correlated through the CKM unitarity.   An unusual pattern of SM predictions results from this study with
  some  existing tensions being dwarfed and new tensions being born.   In particular using HPQCD $B^0_{s,d}-\bar B^0_{s,d}$ hadronic matrix elements a $3.1\sigma$ tension in $\Delta M_s$ independently of $\gamma$ is found.
    For $60^\circ\le\gamma\le 75^\circ$  the tension in $\Delta M_d$ between $4.0\sigma$ and $1.1\sigma$ is found and in the case of $\varepsilon_K$
between $5.2\sigma$ and $2.1\sigma$.
  Moreover, the room for new physics in $K^+\rightarrow\pi^+\nu\bar\nu$,
  $K_L\rightarrow\pi^0\nu\bar\nu$ and $B\to K(K^*)\nu\bar\nu$ decays is significantly increased. We compare the results in this EXCLUSIVE
  scenario with the HYBRID one in which $|V_{cb}|$  in the former scenario
  is replaced by the most recent inclusive $|V_{cb}|$ and  present
  the dependence of all observables considered by us in both scenarios as functions of $\gamma$. As a byproduct we compare the determination of $\vcb$
  from $\Delta M_s$, $\Delta M_d$, $\varepsilon_K$ and $S_{\psi K_S}$ 
  using $B^0_{s,d}-\bar B^0_{s,d}$ hadronic matrix elements from LQCD with
  $2+1+1$ flavours, $2+1$ flavours and their average. Only for the
  $2+1+1$ case values for $\beta$ and $\gamma$ exist for which the
  same value of $\vcb$ is found: {$\vcb=42.6(4)\times 10^{-3}$, 
  $\gamma=64.6(16)^\circ$ } and $\beta=22.2(7)^\circ$. This in turn
  implies a $2.7\sigma$ anomaly in  $B_s\to\mu^+\mu^-$.
\end{abstract}

\thispagestyle{empty}
\newpage
\tableofcontents
\newpage
\setcounter{page}{1}

%
%
%
\section{Introduction}
The rare $K$ and $B$ decays and the quark mixing being GIM \cite{Glashow:1970gm} suppressed in the Standard Model (SM) and simultaneously
being often theoretically clean are very powerful tools for the search of
New Physics (NP) \cite{Buras:2020xsm}.
Unfortunately the persistent tension between inclusive and exclusive determinations of  $|V_{cb}|$ (see e.g.
\cite{Bordone:2019guc,Bordone:2021oof,Ricciardi:2021shl,Aoki:2021kgd}) weakens this power
significantly. As recently reemphasized by us \cite{Buras:2021nns} this is in particular the case   of the branching ratios for rare $K$-meson decays 
  and the parameter $\varepsilon_K$ that exhibit stronger $|V_{cb}|$ dependences than rare $B$ decay branching ratios and the $\Delta M_{s,d}$ mass differences.
  Also similar tensions in the determination of $\vub$ \cite{Leljak:2021vte}  matter.

  One possible solution  to cope with this difficulty is to consider
  within the SM suitable ratios of  two properly chosen observables
  so that the dependences on $\vcb$ and $\vub$ are eliminated
  \cite{Buras:2003td,Bobeth:2021cxm,Buras:2021nns}. While in \cite{Buras:2003td,Bobeth:2021cxm} $B$ physics observables were considered, the analysis in
  \cite{Buras:2021nns} was dominated by the $K$ system and its
  correlation with rare $B$ decays and $B_{s,d}^0-\bar B_{s,d}^0$ mixing.
  In this manner we could construct 16 $\vcb$-independent ratios that
  were either independent of the CKM parameters or only dependent
  on the angles $\beta$ and $\gamma$, that can be determined in tree-level
  processes. Having one day precise experimental values for the ratios
  in question and also precise values on $\beta$ and $\gamma$ will hopefully allow one to identify particular pattern of deviations from SM expectations
  independently of $\vcb$ pointing towards a particular extension of the
  SM.

  But these ratios, even if useful in the context of the tensions in question,
   are not as interesting as
  the observables themselves. Therefore, assuming in addition no
  NP in   $\varepsilon_K$, $\Delta M_d$ and $\Delta M_s$  and in the mixing induced   CP-asymmetry $S_{\psi K_S}$, these ratios allowed to obtain 
  $\vcb$-independent SM predictions for a number
  of branching ratios \cite{Buras:2021nns}. As these four quark mixing observables are very
  precisely measured and theoretically rather clean, the resulting SM predictions
  obtained in this manner turned out to be the most precise to date.
  A brief summary of the results of this analysis just appeared \cite{Buras:2022nrb}.

Another insight in this problematic has been provided recently by the authors
of \cite{Altmannshofer:2021uub} who made a  determination of $\vcb$
and $\vub$ from loop processes, rare decays and quark mixing, by assuming
no NP contributions to these observables. To this end they could use
only well measured observables in the $B$ system and $\varepsilon_K$. This strategy has already been explored in \cite{Buras:2015qea} but there only
$\varepsilon_K$, $\Delta M_d$ and $\Delta M_s$  and $S_{\psi K_S}$ have been considered.

There is no question about that the analyses in \cite{Buras:2003td,Bobeth:2021cxm,Buras:2021nns} will help us to identify possible departures from SM predictions for the $\vcb$-independent ratios and possible pattern of $\vcb$
determinations from various loop processes as analysed in  \cite{Altmannshofer:2021uub,Buras:2015qea}, but also to some extent in \cite{Buras:2021nns}. See in particular
Figs.~12 and 14 of {the latter paper.} Yet, eventually the most obvious
procedure to look for NP is to determine all CKM parameters in tree-level processes under the assumption that NP contributions to these decays
are negligible. This assumption is more likely to be correct than assuming
no NP contributions in loop induced decays. Subsequently the resulting
values of the CKM parameters inserted into SM  amplitudes for  loop induced processes 
would allow for definite predictions for GIM suppressed observables.

 In this spirit in the present paper we follow a more direct but a novel route,
 which to our knowledge has not been explored by anybody to date, at least as far as SM predictions for theoretically clean {observables} {like rare $K$ and $B$ decays, $\Delta M_{s,d}$ and the mixing induced CP asymmetry $S_{\psi K_S}$} are concerned. Instead
 of using the values of $\vcb$ and $\vub$ resulting from the global
 fits of the CKM matrix \cite{Bona:2007vi,Charles:2004jd}, we assume, in contrast to the prevailing {\em inclusive}  expectations
  in the case of $\vcb$,   that the future values of both $\vcb$ and $\vub$ will be determined from {\em exclusive} tree-level  decays.
  Therefore { we use FLAG \cite{Aoki:2021kgd} averages,\footnote{In fact we will use for $\vcb$ its preliminary value that should appear in the 2022 FLAG's edition.} which are based on a number of LQCD calculations that are listed
    after (\ref{FLAGVUB}).}
  With the precisely known $\vus$ the
  resulting rare decay branching ratios, $\varepsilon_K$, $\Delta M_d$, $\Delta M_s$  and $S_{\psi K_S}$ depend than only on the angles $\beta$ and $\gamma$
  in the unitarity triangle that are moreover  correlated through the CKM unitarity.   An unusual pattern of SM predictions results from this study.
  Some present {tensions} are dwarfed and new {tensions} are born. In some cases
  their sizes depend sensitively on the value of $\gamma$ which enhances
  the importance of  precise measurements of this parameter, stressed in
  particular in \cite{Buras:2021nns,Blanke:2018cya}.

  The view that exclusive decays will eventually lead to the best determination of $\vcb$   is rather unusual but has been already expressed by the first author  in the   past \cite{Buras:2020xsm}. The point is that  precise measurements of formfactors by Lattice   QCD (LQCD) accompanied by improved measurements of the relevant branching ratios should   allow eventually a   better control over theoretical uncertainties than it is possible   in inclusive decays and consequently 
  determinations of $\vcb$ and $\vub$ that do not rely on quark-hadron duality.
  Yet, to be on the safe side, in view of the important progress in the inclusive determination of
  $\vcb$ \cite{Bordone:2019guc,Bordone:2021oof,Ricciardi:2021shl,Aoki:2021kgd}, we compare at all stages the results in this EXCLUSIVE
  scenario with the HYBRID one in which $\vcb$  in the former scenario
  is replaced by the most recent inclusive $\vcb$ from \cite{Bordone:2021oof}.

  To our knowledge in the literature only the authors of \cite{Kim:2021rin} performed a similar
  study, but only for $\varepsilon_K$, finding, similar to us, a significant 
  deviation of the SM prediction from the data. However, their analysis
  differs from ours in that for the CKM parameters they used the values
  obtained from global fits of the UT which can be questioned because in fact these SM global analyses used already $\varepsilon_K$ in their fits.
   Moreover, until now they did not incorporate
  the theoretical advances in $\varepsilon_K$ from \cite{Brod:2019rzc} which
  have been taken by us into account in  \cite{Buras:2021nns} and also in the present analysis.

  The outline of our paper is as follows. In Section~\ref{sec:2} we set up
  our strategy as far as CKM parameters are concerned. We also list the
  input parameters used in our numerical analysis. However, we refrain from
  the expressions for the observables which have been studied already by us in
  \cite{Buras:2021nns} and are collected there and in \cite{Buras:2020xsm}.
  The numerical analysis is presented in Section~\ref{sec:3} with 
  the SM predictions for many observables resulting from {the EXCLUSIVE LQCD scenario} and from the HYBRID scenario. {In Section~\ref{HPQCD} we calculate
  the impact of the hadronic matrix elements with $2+1+1$ flavours from the HPQCD collaboration \cite{Dowdall:2019bea} on our results for rare $B$ decays in  \cite{Buras:2021nns}, where the averages of HPQCD results and $2+1$ results from
{Fermilab Lattice and MILC Collaborations (FNAL/MILC)} \cite{Bazavov:2016nty} calculated in \cite{Bobeth:2021cxm} have been used.
We also illustrate how the determination of $\vcb$ from $\Delta M_s$,
    $\Delta M_d$ and $\varepsilon_K$ depends on the number of flavours used in LQCD calculations of the relevant hadronic matrix elements.}     
 We conclude in Section~\ref{sec:4}. Two short appendices list
  LQCD results for  $F_{B_q}$ and  $F_{B_q}\sqrt{\hat{B}_{B_q}}$  for $N_f=2+1$
  and $N_f=2+1+1$.

  \section{Strategy}\label{sec:2}
  The CKM parameters entering our analysis will be 
\be\label{4CKM}
\boxed{\lambda=\vus,\qquad \vcb, \qquad \vub,\qquad \gamma}
\ee
with $\gamma$ one of the  angles in the UT, shown in  Fig.~\ref{UUTa}. It
is equal, within  an excellent accuracy, to the single phase in the standard parametrization of the CKM matrix \cite{Chau:1984fp,Zyla:2020zbs}.

As the input parameters we will use $\lambda=0.225$ and the FLAG
values for $\vcb$ and $\vub$ extracted from the exclusive tree-level decays. Now
these values, as given in the latest FLAG's report, read 
\cite{Aoki:2021kgd} 
\be\label{FLAGVUB}
{\vcb=39.48(68)\times 10^{-3},\qquad 
\vub=3.63(14)\times 10^{-3}, \qquad {(\rm FLAG-2021)}.}
\ee
{These results are based on a number of different LQCD analyses as summarized in Fig.\,38 of \cite{Aoki:2021kgd}. These are from FNAL/MILC \cite{Bailey:2014tva,Lattice:2015tia,Lattice:2015rga,Bazavov:2019aom},
  HPQCD \cite{Na:2015kha,Bouchard:2014ypa,McLean:2019qcx}  and RBC/UKQCD \cite{Flynn:2015mha}  with further details given in the original papers
  and  \cite{Aoki:2021kgd}.}

However, the value for $\vcb$ in (\ref{FLAGVUB}) does not include the most recent one from Fermilab/MILC  \cite{FermilabLattice:2021cdg} that is significantly
lower $38.40(74)\times 10^{-3}$. Fortunately, we were able to obtain from
FLAG a preliminary result for $\vcb$\footnote{We thank Enrico Lunghi for providing this number prior to the official new FLAG's update.} that includes the latter result. Our
basic values for $\vcb$ and $\vub$ obtained from the overall 2022 FLAG's ($\vcb$,$\vub$) fit will be then as follows\footnote{The value
  for $\vcb$ should be considered as preliminary.}
\be\label{FLAGVUB1}
\boxed{\vcb=39.21(62)\times 10^{-3},\qquad 
\vub=3.61(13)\times 10^{-3}, \qquad {(\rm FLAG-2022)},}
\ee
with $\vub$ practically unchanged.
{Larger values for $\vcb$ from exclusive decays using LQCD, in the ballpark  of $41.0\times 10^{-3}$, have been reported in \cite{Martinelli:2021onb,Martinelli:2021myh,Martinelli:2021ccm,Martinelli:2022vvh} and we are looking forward to the 2023
  FLAG report incorporating these results.}

We will also compare our EXCLUSIVE scenario with the HYBRID one
in which the value for $\vcb$ is the inclusive one from \cite{Bordone:2021oof} and the exclusive
one for $\vub$ as above:
\be\label{HYBRID}
\boxed{\vcb=42.16(50)\times 10^{-3},\qquad 
\vub=3.61(13)\times 10^{-3}, \qquad {(\rm HYBRID)}.}
\ee

For  $\gamma$ we will use a broad range $60^\circ\le\gamma\le 75^\circ$. 
Using CKM unitarity the angle $\beta$ in the UT can then be determined
through the correlation of $\beta$ and $\gamma$
\be\label{VTDG}
\boxed{\cot\beta=\frac{1-R_b\cos\gamma}{R_b\sin\gamma}, \qquad
  R_b=(1-\frac{\lambda^2}{2})\frac{1}{\lambda}
\left| \frac{V_{ub}}{V_{cb}} \right|\,.}
\ee
On the other hand  the sides of the UT,
{$R_t$ and $R_b$, can be solely expressed 
in terms of the angles
$\beta$ and $\gamma$, as follows
\cite{Buras:2002yj}
\be\label{RtRb}
R_t=\frac{\sin\gamma}{\sin(\beta+\gamma)}\approx \sin\gamma,
\qquad R_b=\frac{\sin\beta}{\sin(\beta+\gamma)}\approx \sin\beta\,.
\ee
We observe that $R_t$ depends dominantly on $\gamma$, while $R_b$ on $\beta$.
These approximations follow from the experimental fact that
$\beta+\gamma\approx 90^\circ$  and it is an excellent approximation to set $\sin(\beta+\gamma)=1$ in the formulae below although we will not do it in the
numerical evaluations.

The values  of $\vcb$ and $\vub$ in (\ref{FLAGVUB1}) imply
then 
\be
\frac{\vub}{\vcb}=0.0921\pm 0.0036, \qquad R_b=  0.399\pm 0.016 ,\qquad (\text{EXCLUSIVE})
\ee
and {using  \cite{LHCb:2021dcr} 
\be\label{gamma}
  \gamma = (65.4^{+3.8}_{-4.2})^\circ \,.
  \ee
  }
  \be\label{beta}
\beta= (23.50\pm 0.93)^\circ,\qquad S_{\psi K_S}=\sin(2\beta)=0.731\pm 0.024,\qquad(\text{EXCLUSIVE})
\ee
that both differ {mildly} from the measured values \cite{Zyla:2020zbs}
\be\label{betagamma}
  \beta=(22.2\pm 0.7)^\circ, \qquad  S_{\psi K_S}=0.699(17), \qquad (\text{PDG}).
  \ee
  On the other hand in the HYBRID scenario we find
\be
\frac{\vub}{\vcb}=0.0856\pm 0.0032, \qquad R_b=  0.371\pm 0.014,\qquad (\text{HYBRID})
\ee
and 
\be\label{beta1}
\beta= (21.74\pm 0.84)^\circ,\qquad S_{\psi K_S}=\sin(2\beta)=0.688\pm 0.022,(\text{HYBRID})
\ee
in perfect agreement with the experimental measurements   in (\ref{betagamma}).
\begin{figure}
\centering
\includegraphics[width = 0.55\textwidth]{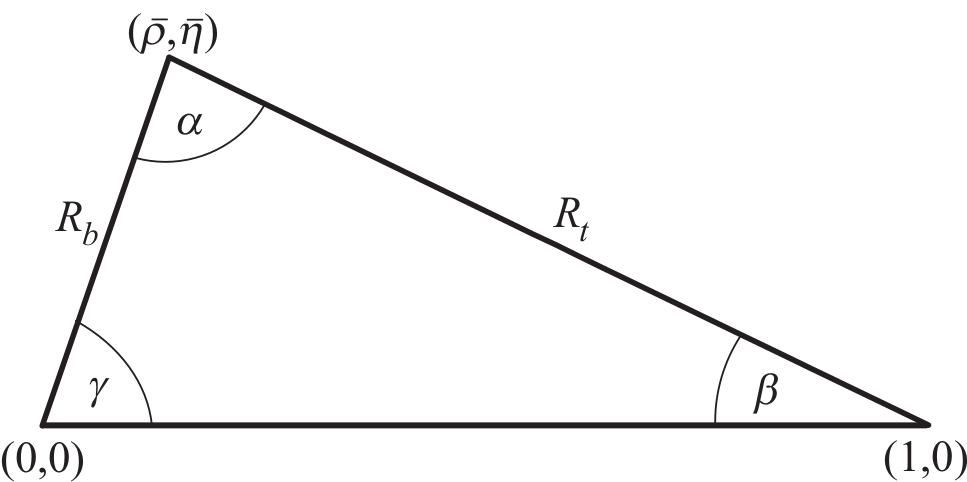}
 \caption{\it The Unitarity Triangle. }\label{UUTa}
\end{figure}

This brief exercise is an overture to the new {tensions} emerging
  from the exclusive strategy.
In Fig.~\ref{Fig:Beta} we show  $\beta$ as a function of $\gamma$ for the  values
of $\vub/\vcb$ in two scenarios in question and compare them to
 the one-sigma range for $\beta$ in (\ref{betagamma}).
We observe that in the case
of the EXCLUSIVE strategy there is indeed a {mild} tension. To remove this tension
a negative NP phase has to be added to $\beta$ in the formula for $ S_{\psi K_S}$
in (\ref{beta}). See (\ref{U21}). This negative phase originates in a NP phase
in $B_d^0-\bar B_d^0$ mixing. {Significantly larger tensions will be
  found in most observables analyzed by us.}

\begin{figure}
\centering
\includegraphics[width = 0.55\textwidth]{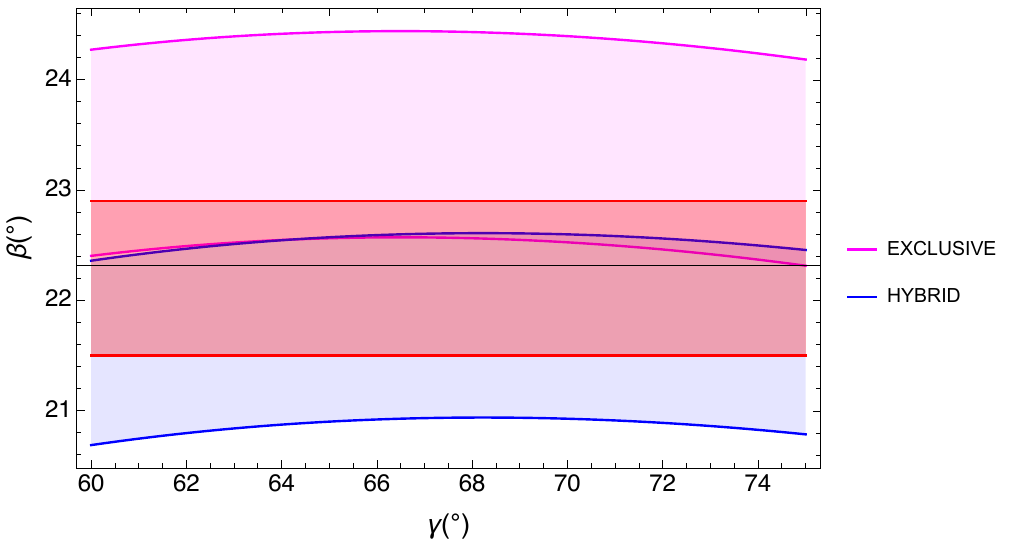}
 \caption{\it The UT angle $\beta$ as  functions of $\gamma$, in the {\rm EXCLUSIVE} and {\rm HYBRID} scenarios. The bands represent the uncertainties related to $\vcb$, $\vub$ and $|V_{us}|$. {The one-sigma range for $\beta$ in (\ref{betagamma}) is  shown as a red band.}}\label{Fig:Beta}
\end{figure}

In this context we would like to mention the analysis in
\cite{Bansal:2021oon} in which the ratio $\vub/\vcb$ was proposed as {a useful
test of the SM}  because of reduced hadronic uncertainties combined with the fact
that this ratio is almost the same for the exclusive and inclusive determinations of $\vcb$ and $\vub$.

Finally, useful are also the following expressions ($\lambda_t=V_{td}V^*_{ts}$)
\be\label{vtdvub}
 \vtd=\lambda  \vcb\sin\gamma=(8.82\pm0.14)\,\sin\gamma \times 10^{-3},\qquad (\text{EXCLUSIVE}),
           \ee
  \be 
  {\IM}\lambda_t=\vub \vcb\sin\gamma=(1.42\pm 0.06)\,\sin\gamma\times 10^{-4},\qquad (\text{EXCLUSIVE}),
\ee
where the values in (\ref{FLAGVUB1}) have been used. For the HYBRID scenario
we have
\be\label{vtdvubH}
 \vtd=\lambda  \vcb\sin\gamma=(9.49\pm0.12)\,\sin\gamma \times 10^{-3},\qquad (\text{HYBRID}),
           \ee
  \be 
  {\IM}\lambda_t=\vub \vcb\sin\gamma=(1.52\pm 0.06)\,\sin\gamma\times 10^{-4},\qquad (\text{HYBRID})\,.
\ee

Moreover
\be\label{vts}
\vts=[1 +\frac{\lambda^2}{2}(1-2 \sin\gamma\cos\beta)]\vcb \approx 0.983\vcb.
\ee

{As $\vtd$ and ${\IM}\lambda_t$ play an important role in rare $K$ and $B$ decays
we show in Fig.~\ref{VtdIM} their dependence on $\gamma$ in both scenarios.}

A recent review of tree-level determinations of $\beta$ and $\gamma$
can be found in Chapter 8 of \cite{Buras:2020xsm}. See also \cite{Descotes-Genon:2017thz,Cerri:2018ypt}. But here we will use $\gamma$ as a free parameter and
$\beta$ as an output by means of (\ref{VTDG}).

\begin{figure}
\centering
\includegraphics[width = 0.55\textwidth]{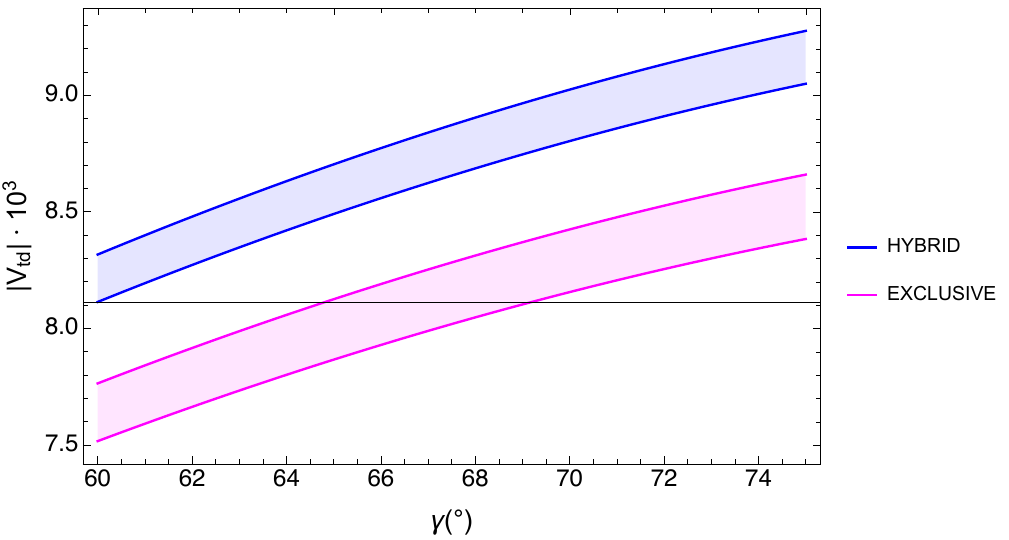}
\includegraphics[width = 0.55\textwidth]{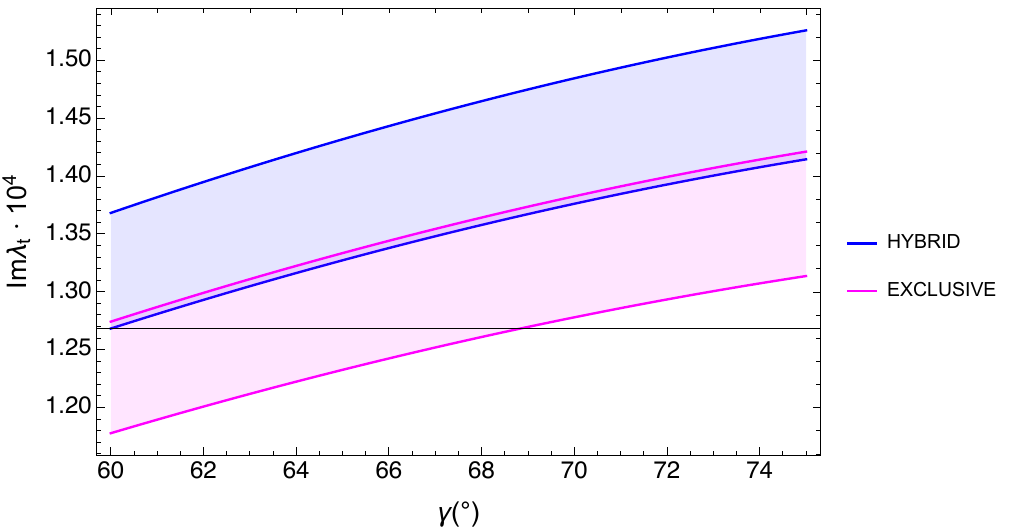}
 \caption{\it $\vtd$ and ${\IM}\lambda_t$ as functions of $\gamma$  in the {\rm EXCLUSIVE} and {\rm HYBRID} scenarios.  }\label{VtdIM}
\end{figure}

\section{Numerical Analysis}\label{sec:3}
Our numerical analysis uses the formulae for various branching ratios that we have
collected in \cite{Buras:2021nns}.
The parameters, other than the CKM ones, entering the formulae in \cite{Buras:2021nns,Buras:2020xsm} are collected in the Table~\ref{tab:input}. Except
for the values of $F_{B_s} \sqrt{\hat B_s}$ and $F_{B_d} \sqrt{\hat B_d}$
that are taken this time {from the HPQCD} collaboration \cite{Dowdall:2019bea}\footnote{These latest LQCD results are in good agreement with the ones from HQET sum rules \cite{King:2019lal}.}, other parameters
are unchanged. These two inputs from $N_f=2+1+1$ LQCD calculations are only
slightly lower than the ones used in \cite{Buras:2021nns,Buras:2020xsm} but are
significantly lower than the ones from $N_f=2+1$ LQCD average given 
by FLAG in  \cite{Aoki:2021kgd}.
These differences are summarized in Appendix~\ref{App:B}. {Their impact on the determination of $\vcb$ from $\Delta M_s$ and $\Delta M_d$ will be analysed in Section~\ref{HPQCD}.}
\begin{table}[!tb]
\center{\begin{tabular}{|l|l|}
\hline
$m_{B_s} = 5366.8(2)\mev$\hfill\cite{Zyla:2020zbs}	&  $m_{B_d}=5279.58(17)\mev$\hfill\cite{Zyla:2020zbs}\\
$\Delta M_s = 17.749(20) \,\text{ps}^{-1}$\hfill \cite{Zyla:2020zbs}	&  $\Delta M_d = 0.5065(19) \,\text{ps}^{-1}$\hfill \cite{Zyla:2020zbs}\\
{$\Delta M_K = 0.005292(9) \,\text{ps}^{-1}$}\hfill \cite{Zyla:2020zbs}	&  {$m_{K^0}=497.61(1)\mev$}\hfill \cite{Zyla:2020zbs}\\
$S_{\psi K_S}= 0.699(17)$\hfill\cite{Zyla:2020zbs}
		&  {$F_K=155.7(3)\mev$\hfill  \cite{Aoki:2019cca}}\\
	$|V_{us}|=0.2253(8)$\hfill\cite{Zyla:2020zbs} &
 $|\eps_K|= 2.228(11)\cdot 10^{-3}$\hfill\cite{Zyla:2020zbs}\\
$F_{B_s}$ = $230.3(1.3)\mev$ \hfill \cite{Aoki:2021kgd} & $F_{B_d}$ = $190.0(1.3)\mev$ \hfill \cite{Aoki:2021kgd}  \\
$F_{B_s} \sqrt{\hat B_s}=256.1(5.7) \mev$\hfill  \cite{Dowdall:2019bea}&
$F_{B_d} \sqrt{\hat B_d}=210.6(5.5) \mev$\hfill  \cite{Dowdall:2019bea}
\\
 $\hat B_s=1.232(53)$\hfill\cite{Dowdall:2019bea}        &
 $\hat B_d=1.222(61)$ \hfill\cite{Dowdall:2019bea}          
\\
{$m_t(m_t)=162.83(67)\GeV$\hfill\cite{Brod:2021hsj} }  & {$m_c(m_c)=1.279(13)\GeV$} \\
{$S_{tt}(x_t)=2.303$} & {$S_{ut}(x_c,x_t)=-1.983\times 10^{-3}$} \\
    $\eta_{tt}=0.55(2)$\hfill\cite{Brod:2019rzc} & $\eta_{ut}= 0.402(5)$\hfill\cite{Brod:2019rzc}\\
$\kappa_\varepsilon = 0.94(2)$\hfill \cite{Buras:2010pza}	&
$\eta_B=0.55(1)$\hfill\cite{Buras:1990fn,Urban:1997gw}\\
$\tau_{B_s}= 1.515(4)\,\text{ps}$\hfill\cite{Amhis:2016xyh} & $\tau_{B_d}= 1.519(4)\,\text{ps}$\hfill\cite{Amhis:2016xyh}   
\\	       
\hline
\end{tabular}  }
\caption {\textit{Values of the experimental and theoretical
    quantities used as input parameters. For future 
updates see FLAG  \cite{Aoki:2021kgd}, PDG \cite{Zyla:2020zbs}  and HFLAV  \cite{Aoki:2019cca}. 
}}
\label{tab:input}
\end{table}

 Here we only recall that  the dependence of the observables considered by us on $\vus$ is negligible. As far as $\beta$ and $\gamma$ are concerned, the angle $\beta$ is already known from the mixing induced CP-asymmetry $S_{\psi K_S}$
 with respectable precision as given in (\ref{betagamma}) and
there is a significant progress by the LHCb collaboration on the determination of $\gamma$ from tree-level strategies \cite{LHCb:2021dcr} so that we have
presently from tree-level decays {the value} given in (\ref{gamma}).
   Moreover, in the coming years the determination of $\gamma$ by the LHCb and Belle II collaborations 
  should be significantly improved so that precision tests of the SM using the
  strategy in \cite{Buras:2021nns} and the one presented here will be possible.

  However we emphasize that we do not use
  the value of $\gamma$ above as an input parameter. Our strategy 
  will be to treat $\gamma$ as a free parameter 
  in the rather broad range $60^\circ\le\gamma\le 75^\circ$ 
  and in view of the future measurements of $\gamma$ by LHCb and Belle II
  to exhibit the $\gamma$ dependence of the observables considered by us.
  We recall that the angle $\beta$ is  the output by means of the unitarity relation (\ref{VTDG}) as given in Fig.~\ref{Fig:Beta}.

  In Table~\ref{tab:SMBRs} we show SM predictions for a number of rare
  $K$ and $B$ branching ratios and $\Delta F=2$ observables resulting from the EXCLUSIVE input in (\ref{FLAGVUB1}) setting $\gamma=65.4^\circ$, the central LHCb
  value. {The uncertainties appearing therein, thus, do not include any error on the $\gamma$ determination.} We also show our results in the HYBRID scenario defined in (\ref{HYBRID}). The latter are not far from the ones obtained in 
  \cite{Buras:2021nns} where the absence of NP contributions to $\Delta F=2$
  observables was assumed. We do not consider the decays like
  $B\to K(K^*)\ell^+\ell^-$ that have larger theoretical uncertainties
  than the observables considered by us. Their $\vcb$ dependence has
  been investigated recently in \cite{Altmannshofer:2021uub}.

   The decay $B_s\to X_s\gamma$ was not considered in \cite{Buras:2021nns}. The result for $B_s\to X_s\gamma$ in both scenarios
  is obtained here from  \cite{Misiak:2015xwa} that effectively corresponds
  to the inclusive $\vcb=42.0\times 10^{-3}$. We just rescaled it using the exclusive and inclusive values of $\vcb$ for EXCLUSIVE and HYBRID scenarios, respectively.

\begin{table}
\centering
\renewcommand{\arraystretch}{1.4}
\resizebox{\columnwidth}{!}{
\begin{tabular}{|l|lll|}
\hline
Decay 
& EXCLUSIVE
& HYBRID
&  DATA
\\
\hline \hline
 $\mathcal{B}(\kpn)\times 10^{11}$ & $6.88(38)$ & $8.44(41)$ & $10.9(38)$\hfill\cite{CortinaGil:2020vlo} 
\\
 $\mathcal{B}(\klpn)\times 10^{11}$ & $2.37(15)$ & $2.74(14)$ & $< 300$ \hfill\cite{Ahn:2018mvc} 
\\
$\mathcal{B}(\ksm)\times 10^{13}$ & $1.49(10) $ & $1.72(8)$ &
$10^4$\hfill\cite{Aaij:2017tia} 
\\
$\overline{\mathcal{B}}(B_s\to\mu^+\mu^-)\times 10^{9}$ & $3.18(12)$ &  $3.67(12)$
&   $2.86(33)$\hfill\cite{LHCb:2021awg,CMS:2020rox,ATLAS:2020acx}
\\
$\mathcal{B}(B_d\to\mu^+\mu^-)\times 10^{10}$ & $0.864(34) $ & $0.999(34)$
&  $<2.05$\hfill\cite{LHCb:2021awg}
\\
$\mathcal{B}(B^+\to K^+\nu\bar\nu)\times 10^{6}$ & $3.83(53)$ & $4.42(60)$ &
  $11\pm 4$\hfill\cite{Browder:2021hbl}
\\
$\mathcal{B}(B^0\to K^{0*}\nu\bar\nu)\times 10^{6}$ & $8.32(82)$ & $9.61(93)$ &
   $<18$\hfill \cite{Grygier:2017tzo}
\\
$\mathcal{B}(B\to X_s\gamma)\times 10^{4}$  &  $2.93(20)$ &    $3.39(23)$  &  $3.32(15)$\hfill \cite{Zyla:2020zbs}
\\
$ |\varepsilon_K|\times 10^{3}$ & $1.78(11)$ &  $2.14(12)$  & $ 2.228(11)$\hfill\cite{Zyla:2020zbs}
\\
$S_{\psi K_S}$  & $0.731(24)$ &  $0.688(22)$  & $0.699(17)$\hfill\cite{Zyla:2020zbs}
\\
 $\Delta M_s \,\text{ps}^{-1}$ &$15.02(87)$ &  $17.35(94)$   & $17.749(20)$\hfill \cite{Zyla:2020zbs}
\\
 $\Delta M_d \,\text{ps}^{-1}$ &$0.434(28)$ &  $ 0.502(31) $  & $ 0.5065(19)$\hfill \cite{Zyla:2020zbs}
\\
\hline
\end{tabular}
}
\renewcommand{\arraystretch}{1.0}
\caption{\label{tab:SMBRs}
  \small
  Predictions (second column) for various observables within the SM using the EXCLUSIVE strategy for $\vcb$ and $\vub$ and $\gamma=65.4^\circ$. In the third column we show the results for the HYBRID choice of $\vcb$ and $\vub$ as given in (\ref{HYBRID}) and in the fourth the experimental data.
}
\end{table}

  In Figs.\ref{Fig:Kp},  \ref{Fig:Bd} and \ref{Fig:epsK}  we show the $\gamma$ dependence of the following observables
  \be
  \mathcal{B}(\kpn),\quad \mathcal{B}(\klpn),\quad \Delta M_d,\quad \mathcal{B}(B_d\to\mu^+\mu^-), \quad \varepsilon_K,\quad  S_{\psi K_S}
  \ee
  in both scenarios for $\vcb$ and $\vub$. $\mathcal{B}(\ksm)$ has the same
  $\gamma$ dependence as $\mathcal{B}(\klpn)$ and the remaining observables
  do not depend {or only weakly depend} on $\gamma$.
  
  \begin{figure}
\centering
\includegraphics[width = 0.55\textwidth]{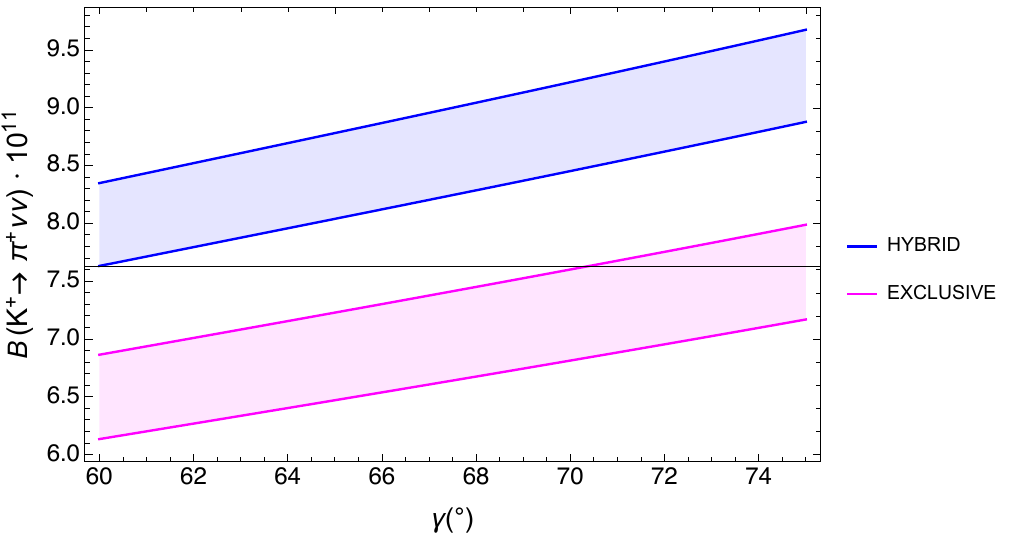}
\includegraphics[width = 0.55\textwidth]{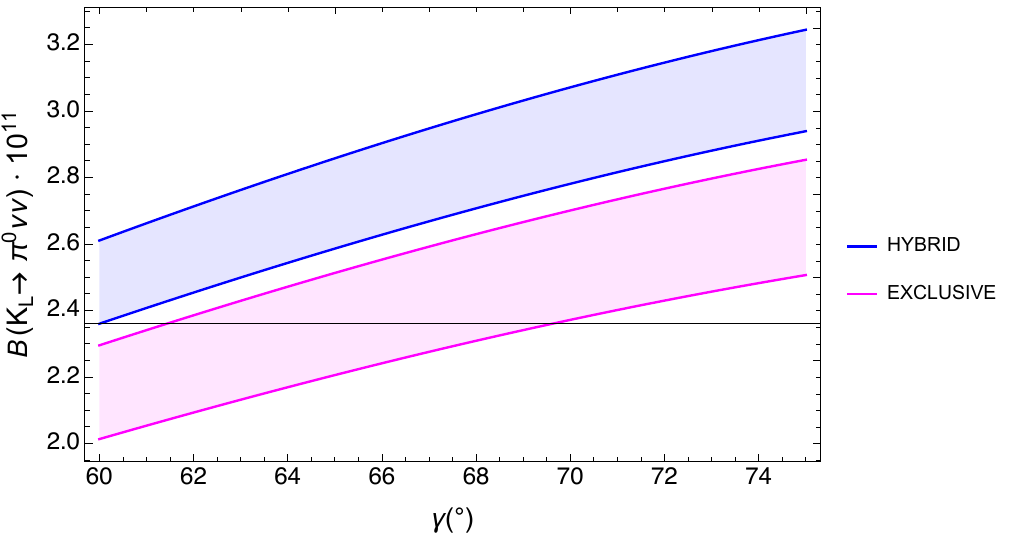}
 \caption{\it The branching ratios $\mathcal{B}(\kpn)$ and $\mathcal{B}(\klpn)$ as  functions of $\gamma$, in the {\rm EXCLUSIVE} and {\rm HYBRID} scenarios. The bands represent the uncertainties related to $\vcb$, $\vub$, $|V_{us}|$ and to the non-CKM parameters. }\label{Fig:Kp}
\end{figure}


 \begin{figure}
   \centering
   \includegraphics[width = 0.55\textwidth]{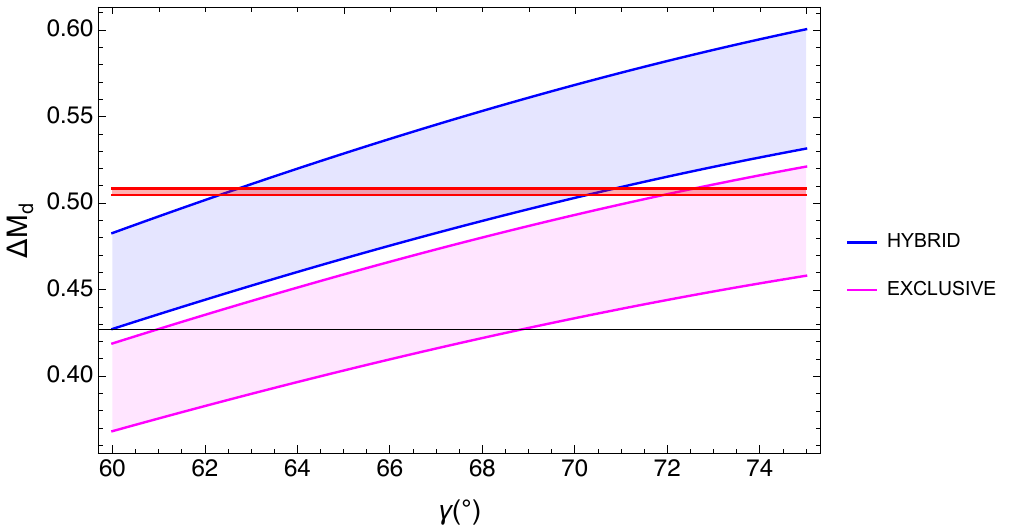}
\includegraphics[width = 0.55\textwidth]{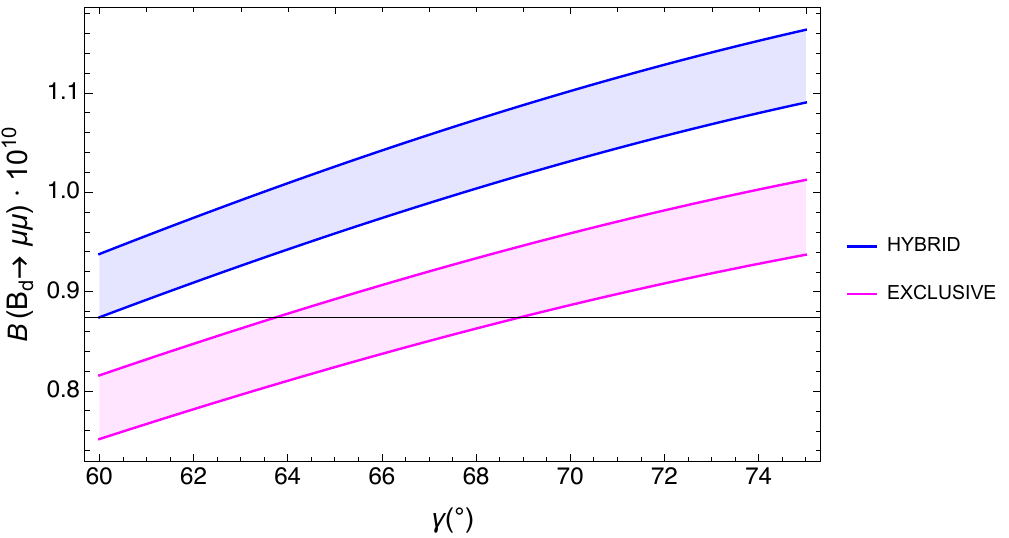}
 \caption{\it $\Delta M_d$  and the branching ratio $\mathcal{B}(B_d\to\mu^+\mu^-)$ as  functions of $\gamma$, in the {\rm EXCLUSIVE} and {\rm HYBRID} scenarios. The bands represent the uncertainties related to $\vcb$, $\vub$, $|V_{us}|$ and to the non-CKM parameters. {The red band in the upper panel represents the experimental value for $\Delta M_d$, with its $1\sigma$ uncertainty.}}\label{Fig:Bd}
\end{figure}

\begin{figure}
\centering
\includegraphics[width = 0.55\textwidth]{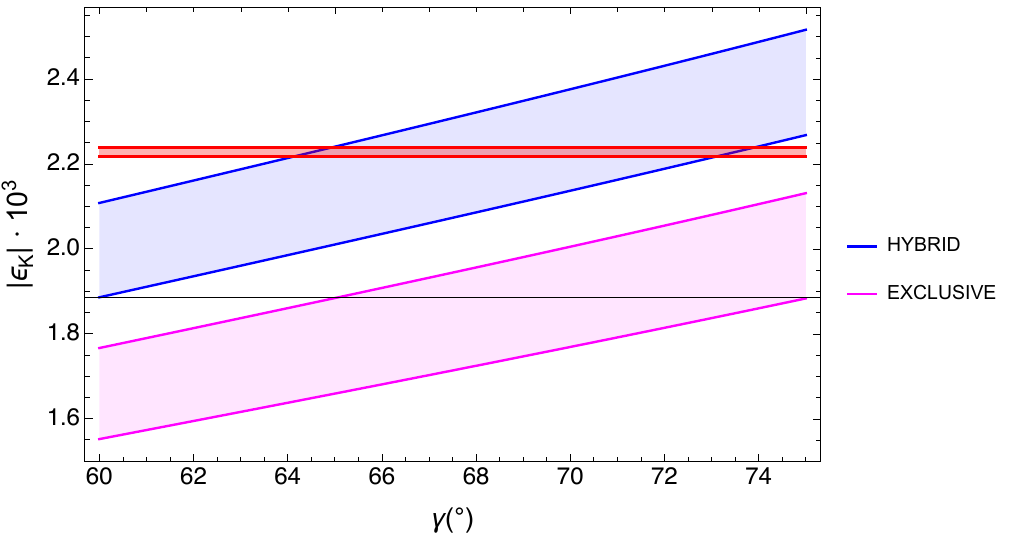}
\includegraphics[width = 0.55\textwidth]{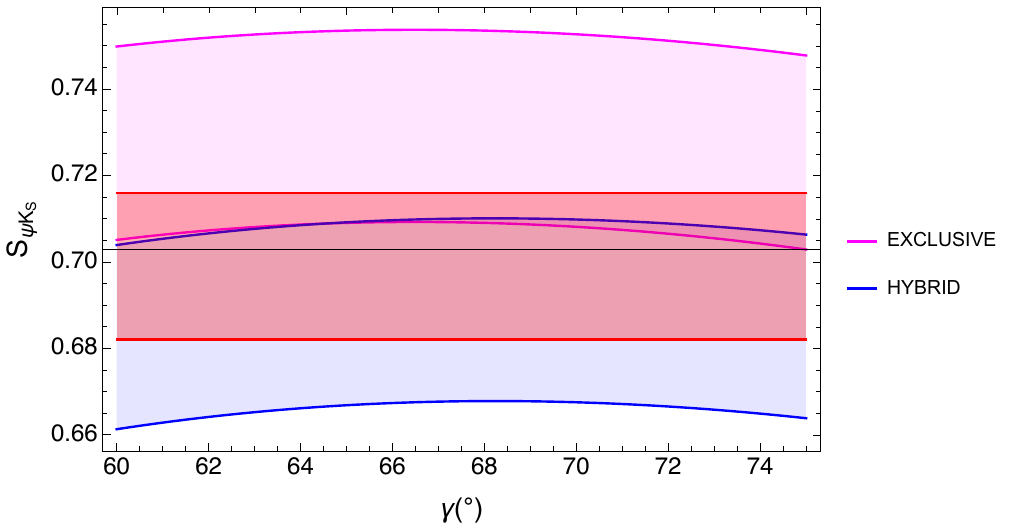}
 \caption{\it $\varepsilon_K$ and the CP asymmetry $S_{\psi K_S}$ as  functions of $\gamma$, in the {\rm EXCLUSIVE} and {\rm HYBRID} scenarios. The bands represent the uncertainties related to $\vcb$, $\vub$, $|V_{us}|$ and to the non-CKM parameters. The red band in the upper panel represents the experimental value for $\varepsilon_K$, with its $1\sigma$ uncertainty. The same for  $S_{\psi K_S}$.}\label{Fig:epsK}
\end{figure}



Concentrating first on the EXCLUSIVE scenario we observe:
  \begin{itemize}
  \item
    As seen in Table~\ref{tab:SMBRs} for $\gamma=65.4^\circ$, the largest tensions
    are found for $\varepsilon_K$ ($4.1\sigma$), $\Delta M_s$ ($3.1\sigma$)
    and $\Delta M_d$ ($2.6 \sigma$).
  \item
    As $\Delta M_s$ is practically independent of $\gamma$ this tension remains for other values of $\gamma$.
   \item
     As seen in Fig.~\ref{Fig:Kp}, the branching ratios for $\kpn$ and $\klpn$
     are significantly suppressed below the values found in the literature
     that are in the ballpark of $8.5\times 10^{-11}$ and $3.0\times 10^{-11}$,
     respectively \cite{Buras:2021nns}. But this suppression decreases with increasing $\gamma$.
   \item
     As seen in Fig.~\ref{Fig:Bd},
     the tension in $\Delta M_d$ decreases
    to ${0.6}\,\sigma$ for $\gamma=75^\circ$ but is as large as $4\sigma$ for
    $\gamma=60^\circ$. The branching ratio for $B_d\to\mu^+\mu^-$ shows
    a similar behaviour because its ratio to $\Delta M_d$ is CKM parameters
    independent. The uncertainty in $\Delta M_d$ is a bit larger because
    of the additional hadronic uncertainty in the parameter $\hat B_d$. {The red band in the upper panel represents the experimental value for $\Delta M_d$, with its $1\sigma$ uncertainty.}
  \item
    As seen in Fig.~\ref{Fig:epsK}, the tension  for $\varepsilon_K$ {(with the experimental measurement shown in red)}  is practically linear in $\gamma$ and in the range of $\gamma$ considered varies
    from ${2.0}\sigma$ for $\gamma=75^\circ$ to $5.2\sigma$ for $\gamma=60^\circ$.
    While significant tension in  $\varepsilon_K$  in the EXCLUSIVE scenario
    has been already identified in \cite{Kim:2021rin}, our analysis differs in several respects from that paper as we already stated at the beginning of this writing.
 On the other hand the tension for $S_{\psi K_S}$ is practically
    independent of $\gamma$ and in the ballpark of $1.0\sigma$ so that
    in this case one really cannot talk about an anomaly.
    \item The tension in $B_s\to\mu^+\mu^-$ basically disappears.
      \end{itemize}

  On the other hand in the HYBRID scenario all these tensions disappear but
  the one in $B_s\to\mu^+\mu^-$ is independently of $\gamma$ in the ballpark
  of $2.1 \sigma$ \cite{Bobeth:2021cxm}.
  
{  
\section{The Impact of the HPQCD Results}\label{HPQCD}
It is of interest to see how the use of $2+1+1$ hadronic matrix elements
from {the HPQCD collaboration} \cite{Dowdall:2019bea} {used in the present paper}, instead of the ones used in \cite{Buras:2021nns} (the average of 2+1 and 2+1+1 results)}, 
would modify our results for rare $B$ decays of the latter paper in which no
NP in $\Delta M_s$ and $\Delta M_d$ has been assumed. We make
this comparison in Table~\ref{tab:SMBRBV1}. For completeness we list there also results
for rare $K$ decays which remain unchanged. This also allows the comparison
with the results obtained in EXCLUSIVE and HYBRID scenarios in Table~\ref{tab:SMBRs}.

{We observe that all $B$ decays branching ratios 
   in  Table~\ref{tab:SMBRBV1} are larger than 
  our results obtained in  \cite{Buras:2021nns} that were rather close to
  the ones in the HYBRID scenario. In view of still sizable experimental
  errors this impact of the HPQCD results cannot be fully appreciated with
  the exception of $B_s\to\mu^+\mu^-$. With the branching ratio for this decay
  in  Table~\ref{tab:SMBRBV1} and the experimental data in Table~\ref{tab:SMBRs}, assuming no NP in $\Delta M_s$, the anomaly   in $B_s\to\mu^+\mu^-$ of $2.1\sigma$ found   in \cite{Bobeth:2021cxm} is raised to $2.7\sigma$.

  {It should be emphasized at this point that all the correlations
    found in   \cite{Buras:2021nns} that do not involve $\Delta M_s$ and
    $\Delta M_d$ remain unchanged but the predictions for $B$ decay branching
    ratios change as we have just seen. This then implies a different SM
    region in the correlation between $\kpn$ and $B_s\to\mu^+\mu^-$ that
    we illustrate in Fig.~\ref{fig:3abis}. There the result
    using  HPQCD $2+1+1$ input (left panel) is compared with the
    one of \cite{Buras:2021nns} (right panel) where the average of $2+1+1$
    and $2+1$ matrix elements from \cite{Bobeth:2021cxm} has been used.
    This difference shows the importance of charm contribution in LQCD
    calculations. Note that the result for $\kpn$ did not change relative
  to  \cite{Buras:2021nns}.}
  
\begin{figure}[t]
\centering%
\includegraphics[width=0.45\textwidth]{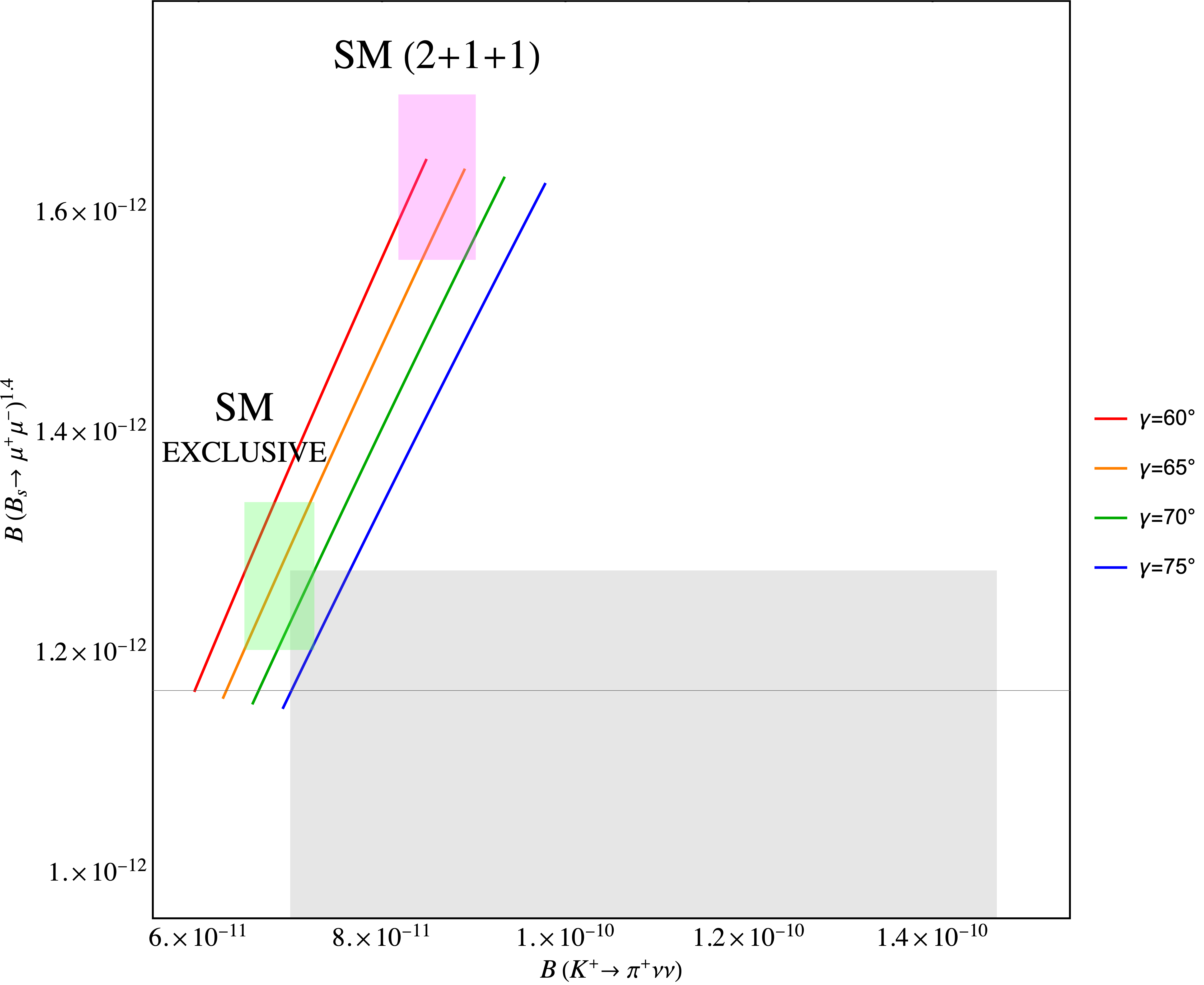}
\includegraphics[width=0.45\textwidth]{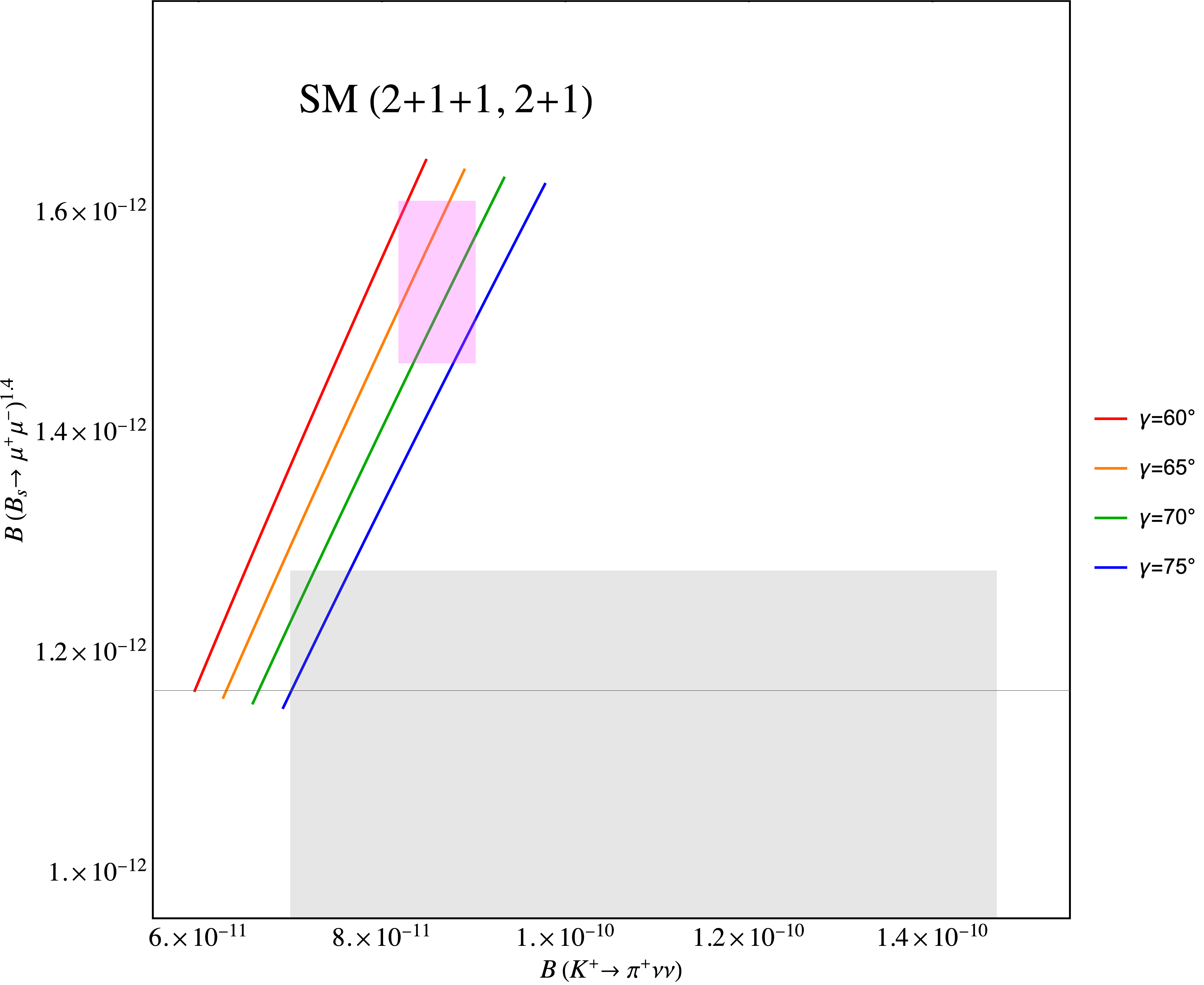}
\caption{\it {The correlation of $\mathcal{B}(\kpn)$ with $\overline{\mathcal{B}}(B_s\to\mu^+\mu^-)^{1.4}$  of  \cite{Buras:2021nns} for different  values of $\gamma$ within the SM. The SM area corresponds to the $HPQCD$ $2+1+1$ input (left panel)     and to the $2+1+1$ and $2+1$ average (right panel) used in  \cite{Buras:2021nns}. The green area represents the EXCLUSIVE scenario and the gray area the
present experimental situation.}} \label{fig:3abis}
\end{figure}

   In fact our results for $B_{s,d}\to\mu^+\mu^-$ are rather consistent
   with the ones obtained by the HPQCD collaboration  \cite{Dowdall:2019bea} in
   2019. But since then the experimental accuracy of $B_{s,d}\to\mu^+\mu^-$
   significantly   increased \cite{LHCb:2021awg,CMS:2020rox,ATLAS:2020acx} allowing a better estimate of the anomaly in question.

  {Finally, we would like to address another important issue.} In \cite{Buras:2021nns}, using the averages of
$B^0_{s,d}-\bar B^0_{s,d}$ hadronic matrix elements from LQCD calculations
with $2+1$ and $2+1+1$ flavours, as given in (\ref{CBAJB}),
we found that there are no values of $\beta$ and $\gamma$ for which
the same value for $\vcb$ can be obtained from  $\varepsilon_K$, $\Delta M_d$ and $\Delta M_s$ when imposing the experimental constraint from $S_{\psi K_S}$.
It is then of interest to investigate what happens when this analysis
is repeated separately for the $2+1$ and $2+1+1$ matrix elements as given
in (\ref{21}) and (\ref{211}), respectively.

The result of this exercise is shown in Fig.~\ref{fig:5}. We observe
that only in the case of $2+1+1$ flavours consistent result for $\vcb$
from all observables considered by us is obtained {which in turn
  provides unique values of $\vcb$ and $\gamma$.
  The determination of $\gamma$ and $\vcb$
can be further improved by considering first the $\vcb$-independent ratio
$\Delta M_d/\Delta M_s$ from which one derives an accurate formula for $\sin\gamma$
\be\label{singamma}
\sin\gamma=\frac{0.983(1)}{\lambda}\sqrt{\frac{m_{B_s}}{m_{B_d}}}\xi
\sqrt{\frac{\Delta M_d}{\Delta M_s}}\,, \qquad \xi=\frac{F_{B_s}\sqrt{\hat{B}_{B_s}}}{F_{B_d}\sqrt{\hat{B}_{B_d}}}=1.216(16),
\ee
with the value for $\xi$ from HPQCD \cite{Dowdall:2019bea} and where (\ref{vts}) has been used. The advantage of
using this ratio for the determination of $\gamma$ over studying $\Delta M_s$ and $\Delta M_d$ separately
is its $\vcb$-independence and the reduced error on $\xi$ from LQCD relative to the individual errors of
hadronic parameters in $\Delta M_s$ and $\Delta M_d$. See Appendix~\ref{App:B}. 
}

{Combining then $\varepsilon_K$, $\Delta M_d$, $\Delta M_s$, $S_{\psi K_S}$
  and using (\ref{singamma}) we obtain finally the following values of
  the CKM parameters
\be\label{CKMR}
\boxed{\vcb=42.6(4)\times 10^{-3},\qquad \gamma=64.6(16)^\circ,\qquad\beta=22.2(7)^\circ.}
\ee
 The value of $\vcb$ is somewhat larger
  than in the HYBRID scenario in (\ref{HYBRID}) but consistent
  with it. It should be noted that the determination of $\gamma$ in this manner is more
  accurate than its present determination from tree-level decays in (\ref{gamma}).}   {The corresponding value of $\vub$ is 
 \be\label{CKMRVub}
\boxed{\vub=3.72(11)\times 10^{-3},}
\ee
which is {slightly} larger than the FLAG determination but consistent with it.} 
  {We observe that $\varepsilon_K$ dominates this determination of $\vcb$.
    This can be traced back to its larger senitivity to $\vcb$ than it is the case for $\Delta M_{s,d}$. While  $\Delta M_{s,d}$ are proportional to $\vcb^2$,
    $|\varepsilon_K|$ exhibits approximately  $\vcb^{3.4}$ dependence \cite{Buras:2021nns}.} Reducing the error on $\beta$, represented by the green band,  and
  decreasing the error on $\gamma$ from its tree-level measurements will
  provide a determination of $\vcb$ with an error below $1\%$.

The $2+1$ case demonstrates significant inconsistencies between $\vcb$ values
from $\Delta M_{d,s}$ and $\varepsilon_K$. The average in (\ref{CBAJB})
considered by us in   \cite{Buras:2021nns} is in a better shape but
also various tensions are identified that we discussed in detail in the latter
paper. The message from this exercise is clear. The inclusion of charm
in the evaluation of $B^0_{s,d}-\bar B^0_{s,d}$ hadronic matrix elements by
LQCD is mandatory and it is important that in addition to HPQCD
\cite{Dowdall:2019bea} a second LQCD collaboration includes charm in the
evaluation of these matrix elements.

Assuming that the HPQCD values will be confirmed by another LQCD
group, the SM predictions in the left panel of Fig.~\ref{fig:3abis}
will be favoured  implying
\be\label{BEST}
\boxed{\overline{\mathcal{B}}(B_s\to\mu^+\mu^-)={(3.78^{+ 0.15}_{-0.10})}\times 10^{-9}, \qquad  \mathcal{B}(\kpn) = (8.60\pm 0.42)\times 10^{-11}.}
\ee

But it should be remembered that in contrast to the EXCLUSIVE scenario
discussed in previous sections these results assume that the SM predictions
for $\Delta M_s$ and $\varepsilon_K$ agree with the data. If the EXCLUSIVE
scenario will turn out to be true, the predictions above will be invalid
and will be replaced by the ones in Table~\ref{tab:SMBRs} {and the green area   in  Fig.~\ref{fig:3abis}.}

\begin{figure}[t!]
  \centering%
  \includegraphics[width=0.70\textwidth]{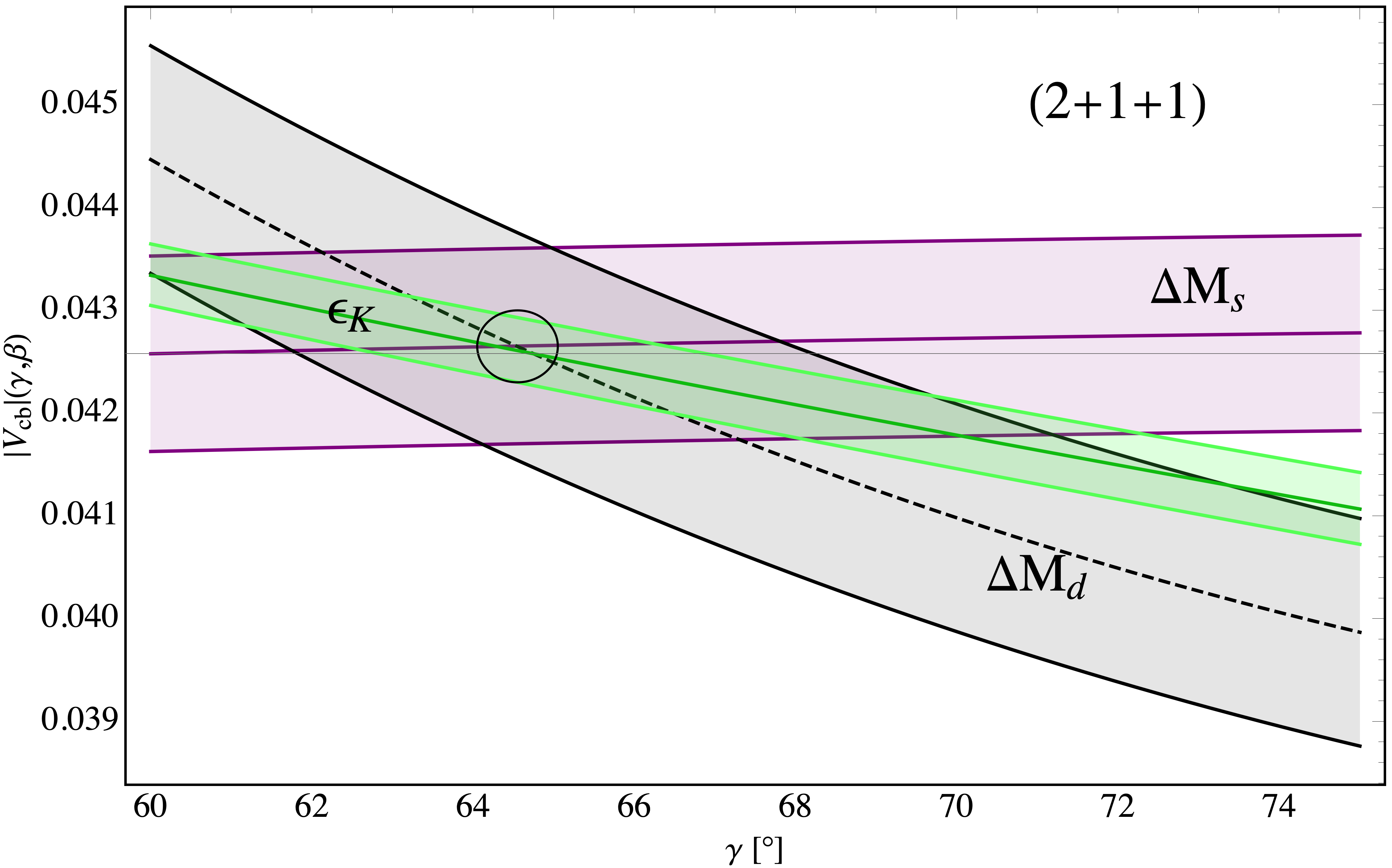}\\
 \includegraphics[width=0.70\textwidth]{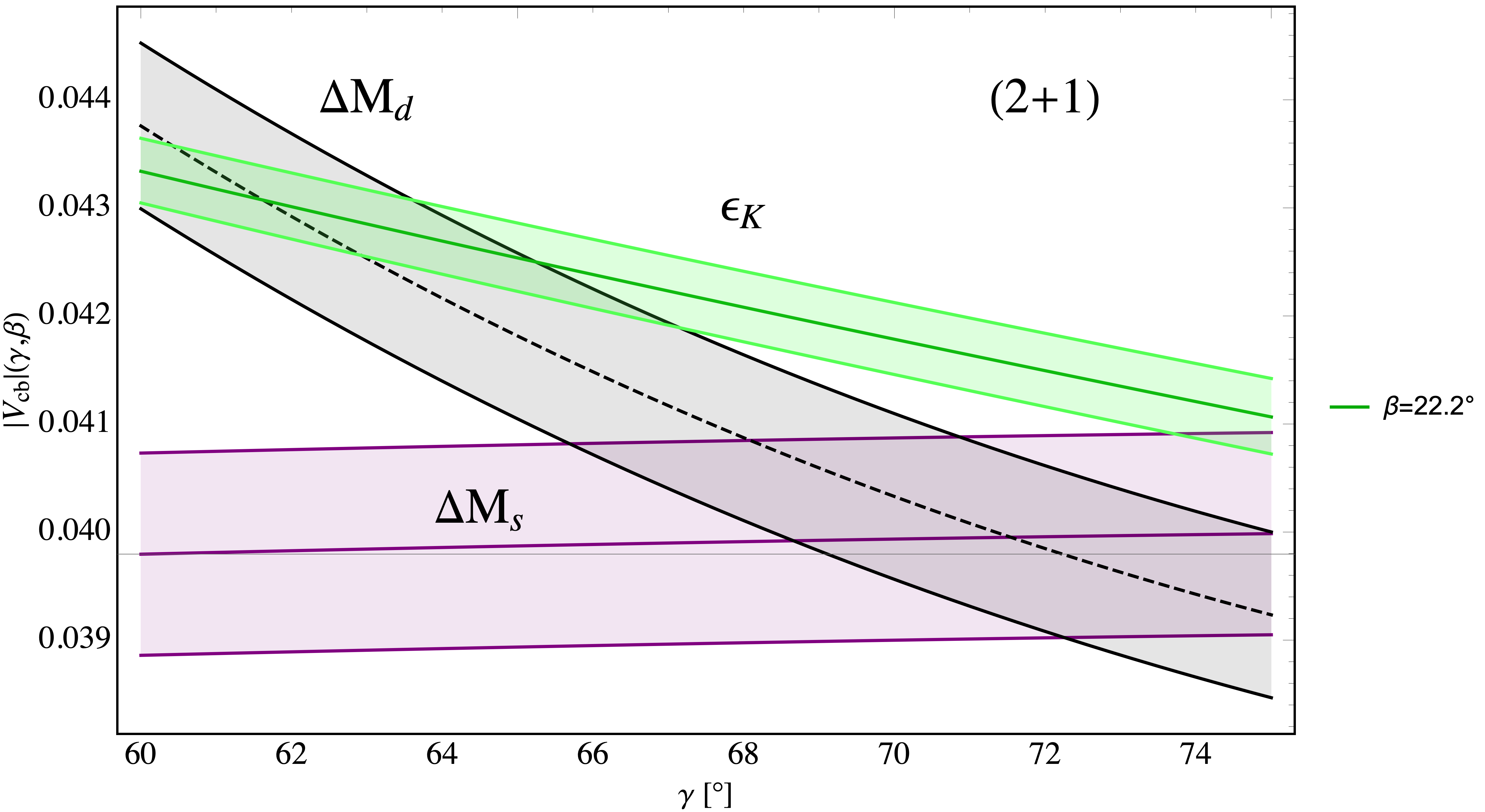}\\
\includegraphics[width=0.70\textwidth]{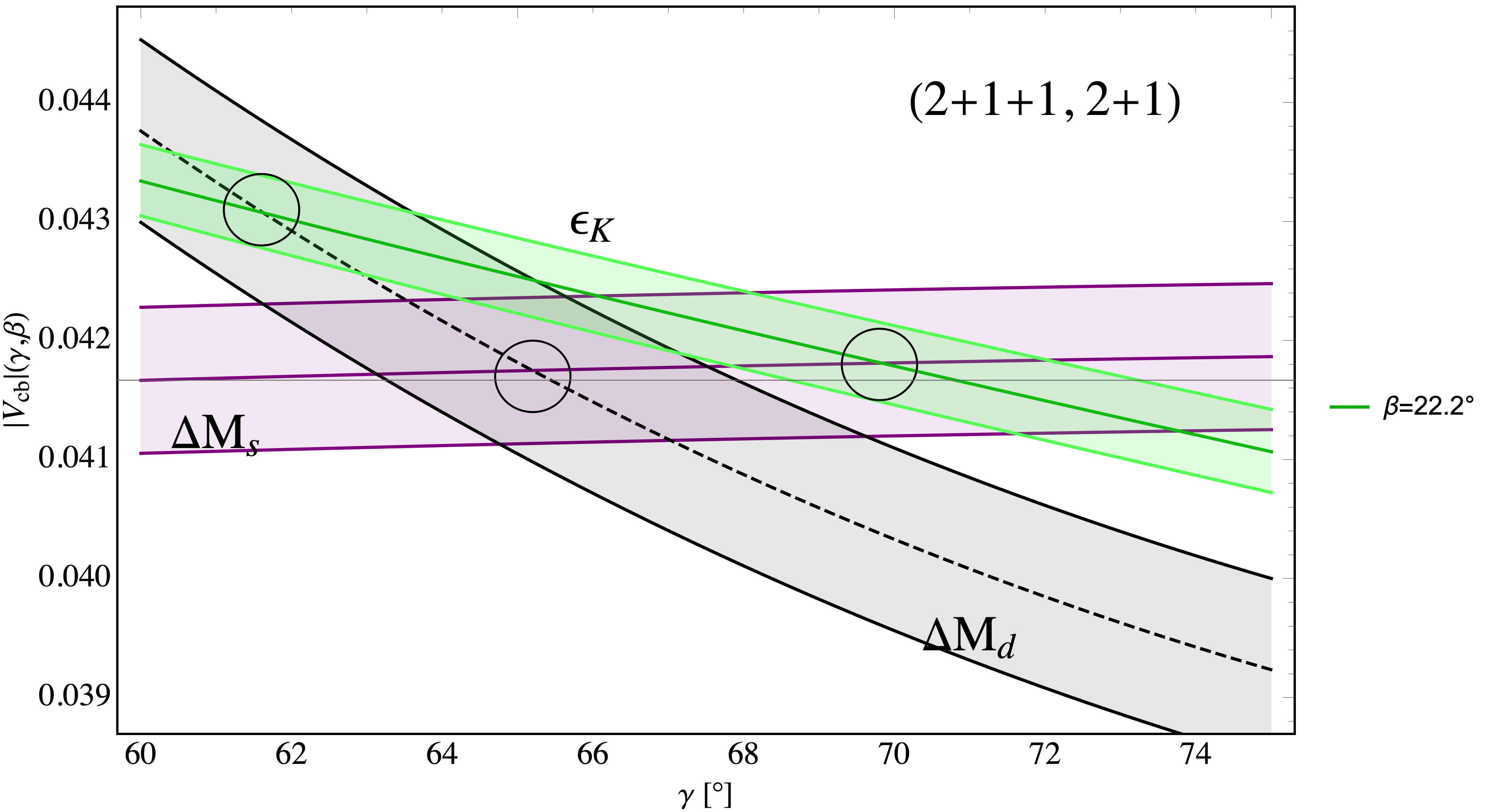}%
\caption{\it {The values of $\vcb$ extracted from $\varepsilon_K$, $\Delta M_d$ and  $\Delta M_s$ as functions of $\gamma$. $2+1+1$ flavours (top), $2+1$ flavours (middle), average of $2+1+1$ and $2+1$
    cases (bottom). The green band represents experimental $S_{\psi K_S}$ constraint on $\beta$.}
\label{fig:5}}
\end{figure}

\begin{table}
\centering
\renewcommand{\arraystretch}{1.4}
\resizebox{\columnwidth}{!}{
\begin{tabular}{|ll|l|}
\hline
Decay 
& Branching Ratio with \cite{Buras:2021nns,Dowdall:2019bea} 
& Branching Ratio with \cite{Buras:2021nns,Bobeth:2021cxm}
\\
\hline \hline
 $B_s\to\mu^+\mu^-$ &  ${(3.78^{+ 0.15}_{-0.10})}\times 10^{-9}$      &    $(3.62^{+ 0.15}_{-0.10})\times 10^{-9}$
\\
 $B_d\to\mu^+\mu^-$ &  ${(1.02^{+ 0.05}_{-0.03})}\ \times 10^{-10}$      & $(0.99^{+ 0.05}_{-0.03})\ \times 10^{-10}$
\\
$B^+\to K^+\nu\bar\nu$ &$ {(4.65\pm 0.62)}\times 10^{-6}$
&    $(4.45\pm 0.62)\times 10^{-6}$
\\
$B^0\to K^{0*}\nu\bar\nu$ & ${(10.13\pm 0.92)}\times 10^{-6}$ & $(9.70\pm 0.92)\times 10^{-6}$
\\
\hline
$\kpn$ & $(8.60\pm 0.42)\times 10^{-11}$ & $(8.60\pm 0.42)\times 10^{-11}$
\\
 $\klpn$ & $(2.94\pm 0.15)\times 10^{-11}$ & $(2.94\pm 0.15)\times 10^{-11}$ 
\\
$\ksm$ & {$(1.85\pm 0.10)\times 10^{-13}$} &  {$(1.85\pm 0.10)\times 10^{-13}$}
\\
\hline
\end{tabular}
}
\renewcommand{\arraystretch}{1.0}
\caption{\label{tab:SMBRBV1}
  \small
  Results for the rare $B$ decay branching ratios using the strategy of 
  \cite{Buras:2021nns} with the $2+1+1$ LQCD hadronic matrix elements of
  \cite{Dowdall:2019bea} (second column) compared with the ones obtained
  in \cite{Buras:2021nns} using the average of $2+1+1$ and $2+1$
  LQCD data from \cite{Bobeth:2021cxm} (third column). Results for rare $K$ decays remain unchanged.}
\end{table}

\section{Conclusions}\label{sec:4}
The EXCLUSIVE vision of rare decays and quark mixing is still not excluded
and could become reality in the coming years. The present paper shows,
similarly to our analysis in \cite{Buras:2021nns}, how important is the determination of $\vcb$ for rare decays, in particular for rare Kaon decays. A precise determination of the $\gamma$ in tree-level decays in the coming years will
shed additional light on the tensions identified by us.

 As we have seen, an unusual pattern of SM predictions results from this study with
  some existing {tensions} disappearing or being dwarfed and new ones being born.
  In particular the $B_s\to\mu^+\mu^-$ tension disappears and instead the anomalies
  at the level of $(2-5)\sigma$ are present in $\Delta M_s$, $\Delta M_d$ and
  in particular in $\varepsilon_K$. While the $3.1\sigma$ tension
  in  $\Delta M_s$ is practically independent of $\gamma$, the one
  in  $\Delta M_d$ increases from $0.6\,\sigma$ to $4\sigma$ when $\gamma$
  is decreased from $75^\circ$ to $60^\circ$. In the case of $\varepsilon_K$ the corresponding variation is from $2.0\sigma$ to $5.2\sigma$.

  Moreover, the room left for NP in $\kpn$,
  $\klpn$ and $B\to K(K^*)\nu\bar\nu$
  is significantly increased but as seen in Figs.\ref{Fig:Kp} and \ref{Fig:Bd} it depends sensitively on $\gamma$.
The tension in $B\to X_s\gamma$ is also interesting.

  It should be recalled that in 2018 with the values of $\gamma\approx 74^\circ$ from the LHCb, with the inclusive $\vcb$ and the $N_f=2+1$ hadronic  $B^0_{s,d}-\bar B^0_{s,d}$ matrix elements, 
  $\Delta M_d$ in the SM was found significantly above the data with a smaller
  enhancement in   $\Delta M_s$ \cite{Blanke:2018cya}. In 2022 with lower values
  for $\gamma\approx 65^\circ$ from the {LHCb \cite{LHCb:2021dcr} }
  and $N_f=2+1+1$ hadronic  $B^0_{s,d}-\bar B^0_{s,d}$ matrix elements from
  {HPQCD} \cite{Dowdall:2019bea} the inclusive values
  of $\vcb$ imply good agreement of the SM with the data on  $\Delta M_{s,d}$.
  But for the exclusive values of $\vcb$ used in the present paper 
  $\Delta M_{s}$ is significantly below the experimental data. This also applies
  to $\Delta M_{d}$ unless $\gamma$ is chosen above $70^\circ$ that is not yet
  excluded by experiments.

  In this context we should emphasize that the $R(K)$ and $R(K^*)$ anomalies
  being independent of $\vcb$ remain. On the other hand, as analysed in
  \cite{Altmannshofer:2021uub}, lowering the value of $\vcb$ decreases the
  anomalies in $B\to K\mu^+\mu^-$, $B\to K^*\mu^+\mu^-$ and $B_s\to\phi\mu^+\mu^-$ decay branching ratios but to remove them  completely values of $\vcb$
  significantly lower than the exclusive ones are  required.

  It is premature to make a detailed analysis of possible BSM scenarios
  that could remove the anomalies in the EXCLUSIVE scenario considered by us. Despite of this let us close
  our paper with a few observations.

  As in the EXCLUSIVE scenario NP is required to enhance  $\Delta M_s$, $\Delta M_d$ and $\varepsilon_K$,
  a natural scenario would be at first sight the constrained Minimal Flavour Violation scenario
  \cite{Buras:2000dm} because, as pointed out in \cite{Blanke:2006yh}, in
  this scenario the $\Delta F=2$ observables can only be enhanced. However,
  the fact that a new phase $\varphi_{\rm new}\approx -1.3^\circ$ is required to fit the data for $S_{\psi K_S}$, 
  a more apprioprate here would be the $\text{U(2)}^3$ scenario
  \cite{Barbieri:2011ci,Barbieri:2011fc,Crivellin:2011fb}.
  As pointed out in \cite{Buras:2012sd}, in the  $\text{U(2)}^3$ scenario the
  CP-asymmetry $S_{\psi K_S}$ is anti-correlated with the CP-asymmetry $S_{\psi\phi}$
\begin{equation}
S_{\psi K_S} = \sin(2\beta+2\varphi_{\rm new})\,, \qquad
S_{\psi\phi} =  \sin(2|\beta_s|-2\varphi_{\rm new})\,,\qquad 
({\rm U(2)^3}),
\label{U21}
\end{equation}
so that with $|\beta_s|\approx 1^\circ$ an  enhancement of the latter asymmetry from the SM prediction $0.0363\pm0.0013$ to $0.080\pm0.020$
would follow. Somewhat above the present data $0.054\pm0.020$  \cite{Zyla:2020zbs} but consistent with it.

As a byproduct we have investigated in Section~\ref{HPQCD} 
  the impact of the hadronic matrix elements from the HPQCD collaboration \cite{Dowdall:2019bea} on our results for rare $B$ decays in  \cite{Buras:2021nns}.
  The most interesting result is the increase of the $B_s\to\mu^+\mu^-$ anomaly from $2.1\sigma$ to $2.7\sigma$.
{Moreover  we compared the determination of $\vcb$
  from $\Delta M_s$, $\Delta M_d$, $\varepsilon_K$ and $S_{\psi K_S}$ 
  using $B^0_{s,d}-\bar B^0_{s,d}$ hadronic matrix elements from LQCD with
  $2+1+1$ flavours, $2+1$ flavours and their average. As seen in Fig.~\ref{fig:5}
only for the
  $2+1+1$ case values for $\beta$ and $\gamma$ can be found for which the
  same value of $\vcb$ is found. The resulting $\vcb$, $\gamma$ and
    $\beta$   are given in (\ref{CKMR}) and $\vub$ in (\ref{CKMRVub}).

 In any case the coming years will hopefully reveal for us which scenario for $\vcb$ and $\vub$ has been chosen by nature. The measurement of $\gamma$
combined with the 16 $\vcb$-independent ratios constructed in
\cite{Buras:2021nns} and with  $\gamma$-dependence of various observables presented here will  also play an important  role in the search for NP.
{The importance of rare $K$ decays in the search for
  NP has been recently summarized in \cite{Aebischer:2022vky} and the prospects
  for reducing hadronic uncertainties in $K$ decays through intensive LQCD computations in the coming years are very good \cite{Blum:2022wsz}.}

The EXCLUSIVE scenario appears to us to be more interesting than the HYBRID one
because it implies more tensions between the SM predictions and the data.
On the other the proponents of the inclusive determinations of $\vcb$
could consider the tensions found by us as an argument against exclusive
determinations of $\vcb$.

\bigskip
  
{\bf Acknowledgements}

\noindent
     We would like to thank Steven Gottlieb and Andreas Kronfeld for the
    discussions on $\vcb$ from LQCD and in particular Enrico Lunghi for providing preliminary value of $\vcb$ from FLAG. {Thanks go also to Fulvia
      De Fazio for a few numerical checks.}
A.J.B acknowledges financial support from the Excellence Cluster ORIGINS,
funded by the Deutsche Forschungsgemeinschaft (DFG, German Research Foundation), 
Excellence Strategy, EXC-2094, 390783311. E.V. has been partially funded by the Deutsche Forschungs-gemeinschaft (DFG, German Research Foundation) under Germany’s Excellence Strategy- EXC-2094 - 390783311, by the Collaborative Research Center SFB1258 and the BMBFgrant  05H18WOCA1  and  thanks  the  Munich  Institute  for  Astro-  and  Particle  Physics(MIAPP) for hospitality.

\appendix
\boldmath
\section{Weak Decay Constants from LQCD}
\unboldmath
For $\Nf=2+1$ FLAG averages {based on \cite{McNeile:2011ng,Bazavov:2011aa,Na:2012sp,Aoki:2014nga,Christ:2014uea,Boyle:2018knm}} read \cite{Aoki:2021kgd} 
\be\label{F1}
F_{B_d} = 192.0(4.3) \;{\rm MeV},\qquad F_{B_{s}} = 228.4(3.7) \; {\rm MeV},
\qquad  \frac{F_{B_{s}}}{F_{B_d}} = 1.201(16)\,,
\ee
while for $\Nf=2+1+1$ FLAG averages {based on \cite{Dowdall:2013tga,Bussone:2016iua,Hughes:2017spc,Bazavov:2017lyh}} are \cite{Aoki:2021kgd} 
\be\label{F2}
F_{B_d} =  190.0(1.3)  \;{\rm MeV},\qquad F_{B_{s}} =230.3(1.3)  \; {\rm MeV},
\qquad  \frac{F_{B_{s}}}{F_{B_d}} = 1.209(5)\,.
\ee
While the central values in (\ref{F1}) and (\ref{F2}) are close to each other,
the latter ones are much more accurate and we use them in our analysis.

\boldmath
\section{Hadronic Matrix Elements from LQCD}\label{App:B}
\unboldmath

For $N_f=2+1$ the FLAG averages dominated by FNAL/MILC results \cite{Bazavov:2016nty} and {including \cite{Gamiz:2009ku,Aoki:2014nga}}are \cite{Aoki:2021kgd} 
\begin{align}\label{21}
        &&  F_{B_d}\sqrt{\hat{B}_{B_d}} &=  225(9) \, {\rm MeV}  
         &  F_{B_s}\sqrt{\hat{B}_{B_s}} &=  274(8) \, {\rm MeV}
	  & \\ 
&&  \hat{B}_{B_d}  &= 1.30(10) 
         &  \hat{B}_{B_s} &=  1.35(6) 
          &\\ 
        &&  \xi  &=  1.206(17) 
        &  \hat{B}_{B_s}/\hat{B}_{B_d}  &=  1.032(38) \,.
          & \label{eq:avxiBB} 
\end{align}

For $N_f=2+1+1$ one finds \cite{Dowdall:2019bea}
\begin{align}\label{211}
    &&   F_{B_d}\sqrt{\hat{B}_{B_d}}&= 210.6(5.5) \;{\rm MeV}\;\;
         & F_{B_s}\sqrt{\hat{B}_{B_s}}&= 256.1(5.7)\; {\rm MeV}
         &\\
      && \hat{B}_{B_d}&= 1.222(61) 
         & \hat{B}_{B_s}&= 1.232(53)\,,
	 &\\
      &&   \xi &=  1.216(16)  
  	&  \hat{B}_{B_s}/\hat{B}_{B_d} & =  1.008(25)\, .
 	& \label{eq:avxiBB4}
\end{align}

In this case there are significant differences between the $N_f=2+1$ and $N_f=2+1+1$ results. Moreover, the latter ones are more accurate and we use them in the
present paper. In  \cite{Buras:2021nns} we have used the averages
of both results given by \cite{Bobeth:2021cxm}
\be\label{CBAJB}
F_{B_d}\sqrt{\hat{B}_{B_d}}= 214.0(39)\,{\rm MeV},\qquad  
F_{B_s}\sqrt{\hat{B}_{B_s}}= 261.7(38)\,{\rm MeV}\,.
\ee
These are consistent with
the ones from  \cite{Dowdall:2019bea} alone but higher.
However FLAG-2021 advices not to make such averages so that this time
we use the $N_f=2+1+1$ values.

%
%
%
\renewcommand{\refname}{R\lowercase{eferences}}

\addcontentsline{toc}{section}{References}

\bibliographystyle{JHEP}

\small

\bibliography{Bookallrefs}

\end{document}